\documentclass[conference]{IEEEtran}
\IEEEoverridecommandlockouts
\usepackage{cite}
\usepackage{amsmath,amssymb,amsfonts}
\usepackage{algorithmic}
\usepackage{graphicx}
\usepackage{textcomp}
\usepackage{booktabs} 
\usepackage{wrapfig}
\usepackage{xcolor}

\begin{document}

\title{Linky: Visualizing User Identity Linkage Results For Multiple Online Social Networks}


\author{\IEEEauthorblockN{Roy Ka-Wei Lee\IEEEauthorrefmark{1},
Ming Shan Hee\IEEEauthorrefmark{2}, Philips Kokoh Prasetyo\IEEEauthorrefmark{3} and
Ee-Peng Lim\IEEEauthorrefmark{4}}
\IEEEauthorblockA{Living Analytics Research Centre\\ 
Singapore Management University\\
Email: \IEEEauthorrefmark{1}roylee.2013@smu.edu.sg,
\IEEEauthorrefmark{2}mshee.2017@smu.edu.sg,
\IEEEauthorrefmark{3}pprasetyo@smu.edu.sg,
\IEEEauthorrefmark{4}eplim@smu.edu.sg}}

\maketitle

\begin{abstract}
User identity linkage across online social networks is an emerging research topic that has attracted attention in recent years. Many user identity linkage methods have been proposed so far and most of them utilize user profile, content and network information to determine if two social media accounts belong to the same person. In most cases, user identity linkage methods are evaluated by performing some prediction tasks with the results presented using some overall accuracy measures. However, the methods are rarely compared at the individual user level where a predicted matched (or linked) pair of user identities from different online social networks can be visually compared in terms of user profile (e.g. username), content and network information. Such a comparison is critical to determine the relative strengths and weaknesses of each method. In this work, we present \textsf{Linky}, a visual analytical tool which extracts the results from different user identity linkage methods performed on multiple online social networks and visualizes the user profiles, content and ego networks of the linked user identities. \textsf{Linky} is designed to help researchers to (a) inspect the linked user identities at the individual user level, (b) compare results returned by different user linkage methods, and (c) provide a preliminary empirical understanding on which aspects of the user identities, e.g. profile, content or network, contributed to the user identity linkage results.
\end{abstract}


\section{Introduction}
\label{sec:introduction}

\textbf{Motivation.} Increasingly, many users find themselves using multiple online social networks (OSNs) such as Twitter, Instagram, Facebook, etc., \cite{pew2014}. These OSNs offer a wide range of functions which capture the users' interests and relationships. As such, one would have to consider the multiple identities of the same user in order to learn the user's interests and social groupings correctly. A better profiling of users will, in turn, improve content and link recommendations in OSNs \cite{carmagnola2009,yan2013}. As such, user identity linkage problem involving multiple OSNs becomes very important.  

In recent years, researchers have proposed several user identity linkage methods and most of these require user profile (e.g. username), content and network information \cite{shu2017}. The existing user identity linkage methods usually return results as a list of matched user identity pairs. Their accuracies are presented in some aggregated measures. However, very few of these methods have been evaluated at the individual user level where the pair of linked user identities are examined in terms of user profile, content and network information.  This will help us determine the strengths and weaknesses of each method.

\textbf{Objectives.}  In this paper, we seek to address the above limitation by proposing \textsf{Linky}\footnote{Website URL temporarily hidden for triple blind.}, an interactive visual analytical tool, which allows users to explore and inspect the linked user OSN identities belonging or predicted to belong to a specific user by various existing user identity linkage methods. The goal of this proposed tool is to enable researchers to (a) inspect the linked user identities at the individual user level, (b) compare results returned by different user linkage methods, and (c) provide a preliminary empirical understanding on which aspects of the user identities, e.g. profile, content or network, contributed to the user identity linkage results.

\begin{figure*}[t]
	\begin{center}
		\includegraphics[scale = 0.33]{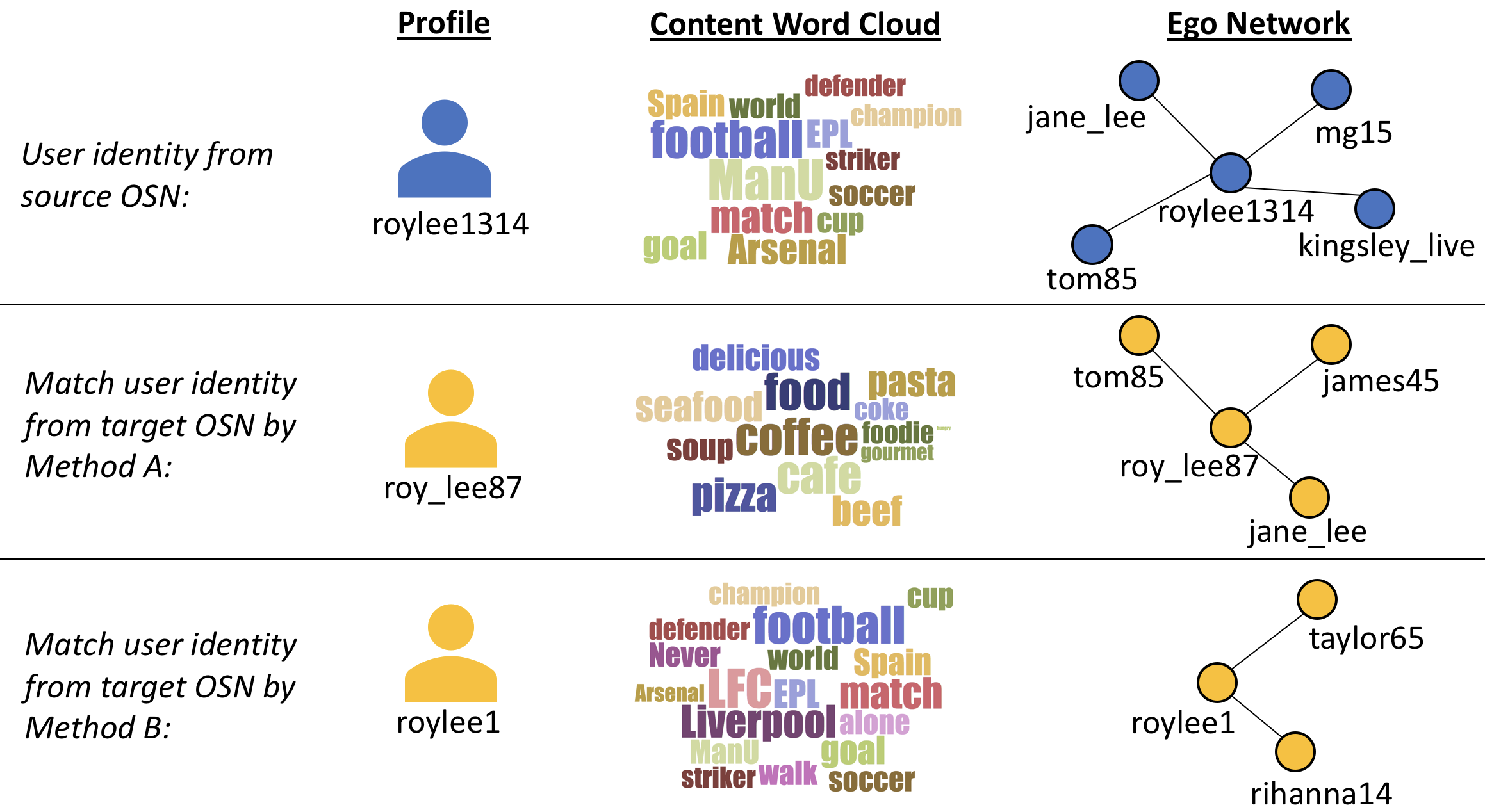}
		\caption{An example of user identity linkage}
		\label{fig:example}
	\end{center}
\end{figure*}

\textbf{Contributions.} The contributions of our demonstration are two-fold. Firstly, it provides researchers an interactive platform to compare user identity linkage results at the individual user level. Currently, most existing user identity linkage methods are evaluated and benchmarked against one other by performing some prediction tasks. For example, a prediction task may be defined as for each user identity $u^s$ in a source OSN, predict the corresponding user identity $u^t$ in a target OSN such that both user identities are likely to belong to the same user. Based on the prediction results return by a user identity linkage method, accuracy metrics such as \textit{precision}, \textit{recall}, \textit{F1} scores, etc., can be derived. To complement these aggregated metrics, \textsf{Linky} enables researchers to examine and compare the visualized user profile attributes, ego networks and generated content of each pair of user identities returned by different user identity linkage methods. In this way, researchers could gain micro-level insights of the user identity linkage results, and determine the strengths and weaknesses of each user identity linkage method. Figure \ref{fig:example} shows an example of user identity linkage where, given a source OSN user identity \textit{roylee1314}, user identity linkage \textit{method A} and \textit{method B} return the corresponding target OSN user identity \textit{roy\_lee87} and \textit{roylee1} respectively. Examining into the user profile attributes, ego networks and generated content of the user identity, we observe that \textit{method A} utilizes more of the network information in matching the user identities whereas method B assigns more weight to the content features.  

Secondly, our demonstration provides a visual preliminary empirical understanding on which aspects of the user identities, e.g. profile, content or network, contribute to the user identity linkage results. Past research have shown that users can exhibit different interests \cite{lee2017}, generate different content \cite{ottoni2014,lim2015} and maintain different relationship networks \cite{lee2016} in different OSNs making user identity linkage very challenging. \textsf{Linky}'s interactive interface will permit researchers to determine such cases and to explore novel solutions to these cases (which unfortunately are beyond the scope of this demonstration). Referencing the same example in Figure \ref{fig:example}, suppose the target OSN user identity \textit{roy\_lee87} returned by method A is the correct match base on ground truth, we might infer that the network information either is more important than content in linking identities for this type of users or generally performs better in this dataset.

\textbf{Paper Outline.} The rest of this paper is organized as follows. We first define the user linkage problem and give an overview of the existing user linkage methods in Section~\ref{sec:linkage}. We present the \textsf{Linky}'s system architecture in Section~\ref{sec:architecture} and describe our demonstration scenario in Section~\ref{sec:demo}. Finally, we conclude our paper in Section~\ref{sec:conclusion}.        

\section{user Identity linkage}
\label{sec:linkage}

User identity linkage for multiple OSNs is a well-studied problem that comes with different formulations \cite{liu2014,mu2016,gao2015,shen2014,kong2013,buccafurri2012,zafarani2013connecting}. Underlying these formulations is the common purpose of matching user identities from two different OSNs. Typically, given two given OSNs $G^s$ and $G^t$, we predict among the given user identity pairs the ones that belong to the same real persons, i.e., $\mathcal{F}:U^s \times U^t \rightarrow \{0,1\}$ such that
\begin{equation}
\begin{aligned}
\mathcal{F}(u^s, u^t) &= \begin{cases}
1 &\textit{if $u^s$ and $u^t$ belong to the same person}\\
0 &\textit{otherwise}
\end{cases} \\
& \textit{where $\mathcal{F}$ is a prediction function to be learnt}
\end{aligned}
\end{equation}

Many methods have been proposed to address the user identity linkage problem. Shu et al. comprehensively summarized these methods in their recent survey \cite{shu2017}. In the survey, they generalized the existing methods into a unified framework which consists of two main phases: \textit{(i)} feature extraction and \textit{(ii)} model construction. In the feature extraction phase, features such as user attributes (e.g., username), generated content and relationship networks are extracted for users from the source and target OSNs. The extracted features are subsequently used as input for the model construction phase, where a supervised, semi-supervised or unsupervised model is trained. Finally, the trained model is used to predict if any two user identities can be linked. 

\textbf{Baseline.} For this demonstration, we adopt the above unified framework and include a simple baseline user identity linkage method which utilizes the username attribute. Username, which is often an alphanumeric string that uniquely identifies an individual user in an OSN, is a minimum common user attributes available on all OSNs. Past research has reported that 59\% of users prefer to use the same username repeatedly on different OSNs for easy recall \cite{zafarani2013connecting}, thus making username a strong and important feature for user identity linkage. 

In our proposed baseline, we first build an n-gram inverted index using all usernames from source and target OSNs. Each source OSN username is then represented as a sparse vector, $\theta_s$, with each vector element corresponding to an n-gram.  The vector element value is the product of occurrence count of the corresponding n-gram in the username and the inverse name frequency of the n-gram. The latter is defined by the reciprocal of the number of names containing the n-gram.  Each username in target OSN is similarly represented as a username vector, $\theta_t$. Cosine similarity is then applied on the username vectors of a pair of source and target OSNs user identities to determine how likely they belong the same user (Refer to Eqn \ref{eqn:cossim}). For a given user identity from the source OSN, the target OSN user identities are ranked by decreasing cosine similarity. In this demonstration, we have used a 3-gram to construct $\theta_s$ and $\theta_t$.

\begin{equation*}
\begin{aligned}
\label{eqn:cossim}
\mathcal{F}(u^s, u^t) &\approx \\
sim(\theta_s,\theta_t) &= \frac{\theta_s \cdot \theta_t}{||\theta_s||||\theta_t||} = \frac{\sum^n_{i=1}(\theta_{si} \theta_{ti})}{\sqrt{\sum^n_{i=1}(\theta_{si})^2} \sqrt{\sum^n_{i=1}(\theta_{ti})^2}}
\end{aligned}
\end{equation*}

\section{System Architecture}
\label{sec:architecture}

\begin{figure} [h]
	\begin{center}
		\includegraphics[scale = 0.9]{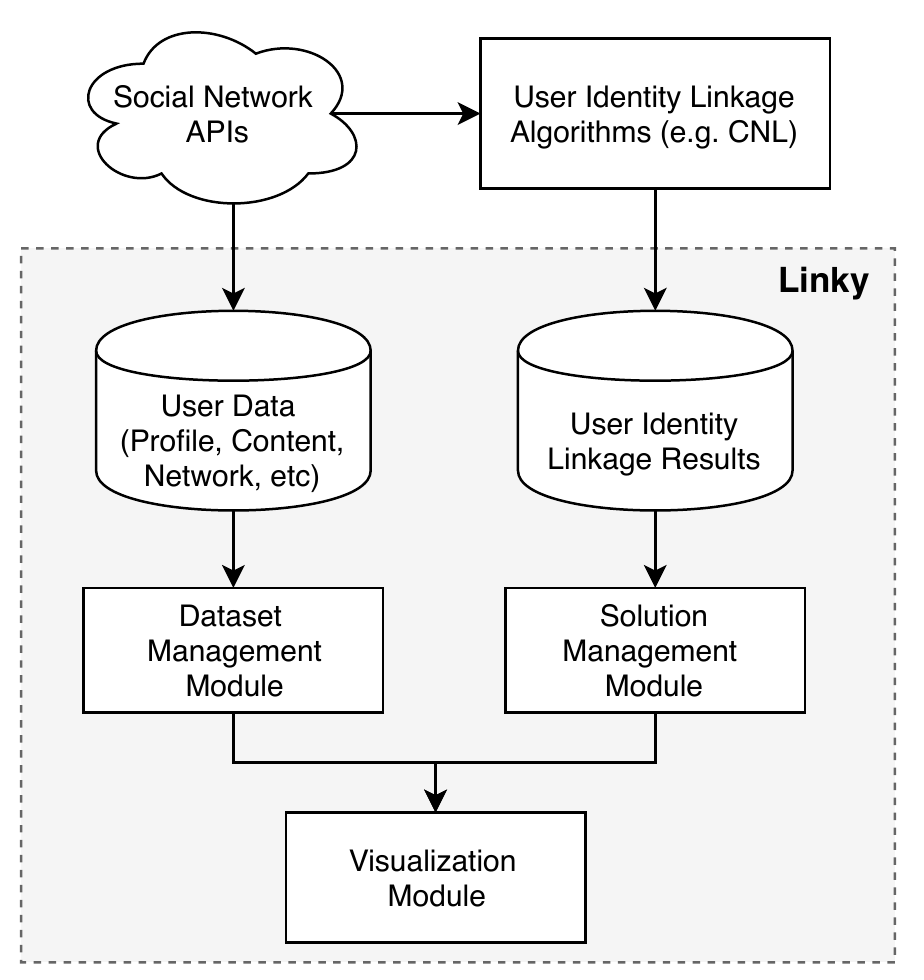}
		\caption{Linky System Architecture}
		\label{fig:Architecture}
	\end{center}
\end{figure}

Figure~\ref{fig:Architecture} shows the system architecture of \textsf{Linky}. We begin with data collection from multiple OSNs using the OSN provided APIs. The collected user datasets are used as input for user identity linkage algorithms to learn and generate the user identity linkage results. The user identity linkage results will then be uploaded to \textsf{Linky} and managed by the \textit{Solution Management} module. The collected user datasets are also stored and managed by the \textit{Dataset Management} module in \textsf{Linky}. Finally, combining the collected user datasets and user identity linkage results, the \textit{Visualization} module visualizes the user profiles, generated content and ego networks of each predicted user identity pair for analysis.    

\textbf{Dataset Management.} The \textsf{Linky} platform enables researchers to upload and manage their own private datasets for analysis. For this demonstration, we gathered a set of over 200k Singapore-based Twitter users who declared Singapore location in their user profiles. These users were identified by an iterative snowball sampling process starting from a small seed set of well known Singapore Twitter users, followed by traversing the follow links to other Singapore Twitter users until the sampling iteration did not get any more new users. From the Twitter users, we obtained a subset of them who have declared their Foursquare accounts in their short bio descriptions (known as the \textit{cross-platform self-matched user accounts}). Next, we traversed the follow links of the \textit{cross-platform users} to other Foursquare users using the OSN's APIs. As some of these Foursquare users may also mention their Twitter accounts in their Foursquare short bio descriptions, we also gathered the mentioned Twitter accounts. Besides the self-declared matching user accounts, the final dataset could have matched user accounts that are not declared in users' short bios descriptions. Note that the self-declared user accounts are also used as ground truth in prediction tasks performed by the individual user identity linkage methods. Finally, all collected user datasets are stored in ElasticSearch\footnote{https://www.elastic.co/}, a distributed, RESTful search engine based on Lucene. 

\textbf{Solution Management.} \textsf{Linky} allows researchers to upload results from multiple user identity linkage methods for benchmarking and analysis. In this demonstration, besides the baseline described in Section \ref{sec:linkage}, we have also selected two state-of-the-art user identity linkage methods to match user identities using the data that we have collected. The two methods are \textit{ULink} \cite{mu2016}, a supervised model that utilizes a projection algorithm and features such as users' profile attributes and generated content to match OSN user identities, and \textit{CNL} \cite{gao2015}, a semi-supervised model that can handle heterogeneous user profile attributes and missing information in user identity linkage.

\begin{table}[h]
\centering
\caption{Features used in user identity linkage algorithms}
\label{tbl:features}
\begin{tabular}{|l|p{5cm}|}
\hline
\textbf{Feature}              & \textbf{Description}                                                                                                                                                                                             \\ \hline\hline
\textit{Username}             & Username of the account (e.g., roylee87).                                                                                                                                                                        \\ \hline
\textit{Screen name}          & Natural name of the user account. It is usually formed using the first and last name of the user (e.g., Roy Lee).                                                                                                \\ \hline
\textit{Bio description}      & short description about the user provided by the account.                                                                                                                                                        \\ \hline
\textit{Relationship network} & Ego-network of the user account. Note that Facebook has undirected friend relationships, while Twitter, Instagram and Foursquare have directed following relationships.                                          \\ \hline
\textit{Content}              & Content from posts published by the user account is summarized as a word cloud. Note that if the content is an rich media post (e.g., image), we will use the caption or description of the post as its content. \\ \hline
\end{tabular}
\end{table}

\begin{figure}[h]
	\begin{center}
		\includegraphics[scale = 0.37]{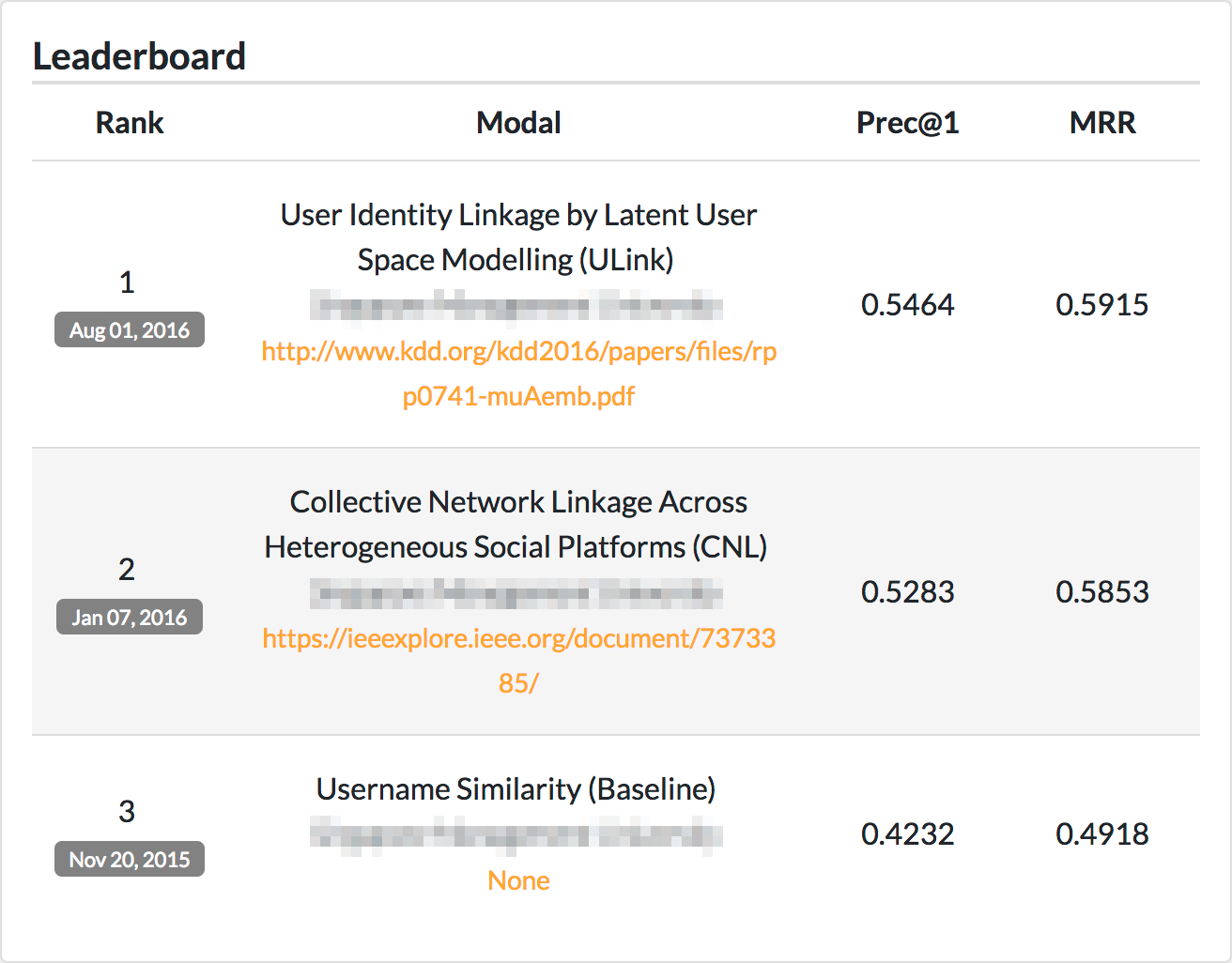}
		\caption{Solution Management Module: List of algorithm uploaded by Linky user}
		\label{fig:solution}
	\end{center}
\end{figure}

We first extracted the features needed to train the baseline and two user identity linkage algorithms from the collected user datasets. Table \ref{tbl:features} shows a list of features extracted. After running the user identity linkage algorithms on the extracted features, the algorithms generate the user identity linkage results where, for each user identity in the source OSN, the algorithms return a ranked-list matching user identity from the target OSN. The user identity linkage results generated by the two algorithms are then uploaded to \textsf{Linky} using the \textit{Solution Management} module. The \textit{Solution Management} module will evaluate the accuracy of the user identity linkage algorithms using ground truth data. Figure \ref{fig:solution} shows the user identity linkage results uploaded to \textsf{Linky} and the overall performance measured (based on ground truth) using aggregated metrics such Precision at top 1 (Prec@1) and Mean Reciprocal Rank (MRR) \cite{Hersh99trec}. User of the \textsf{Linky} platform can click on the results from different algorithms to view the individual user identity pairs matched by the algorithm. 

\textbf{Visualization.} The visualization module is implemented via HTML and JavaScript with Flask\footnote{https://http://flask.pocoo.org/} as the server framework. To query a user and his/her matching user profiles, the user entities, content and ego network information are retrieved from the ElasticSearch engine using the ids of the corresponding user identities. The visualization has of three main components:

\begin{enumerate}
\item \textit{User attributes}. The user attributes (e.g., profile image, username, bio, etc.) are displayed at the top panel of the two matching user profile.
\item \textit{User content word cloud}. The generated content of the user profiles is summarized and displayed as word clouds \cite{cui2010} in the middle panels. A word cloud is an effective visual representation for a collection of words with the frequency of a word shown by its font size, e.g., the bigger the word, the higher frequency it has in the user content. The generation of the word cloud is handled using D3.js \footnote{https://d3js.org/} Javascript library. 
\item \textit{User ego network}. The user ego network is visualized in the bottom panels. To visualize user's ego networks in a meaningful way, we utilize the hierarchical edge bundling algorithm~\cite{holten2006}. Hierarchical edge bundling displays nodes in a circular layout ordering based on their node degrees, thereby reducing clutter in the layout. The hierarchical edge bundling visualization of user's ego network is implemented using the D3.js Javascript library. 
\end{enumerate}

\begin{figure*}[t]
	\begin{center}
		\includegraphics[scale = 0.35]{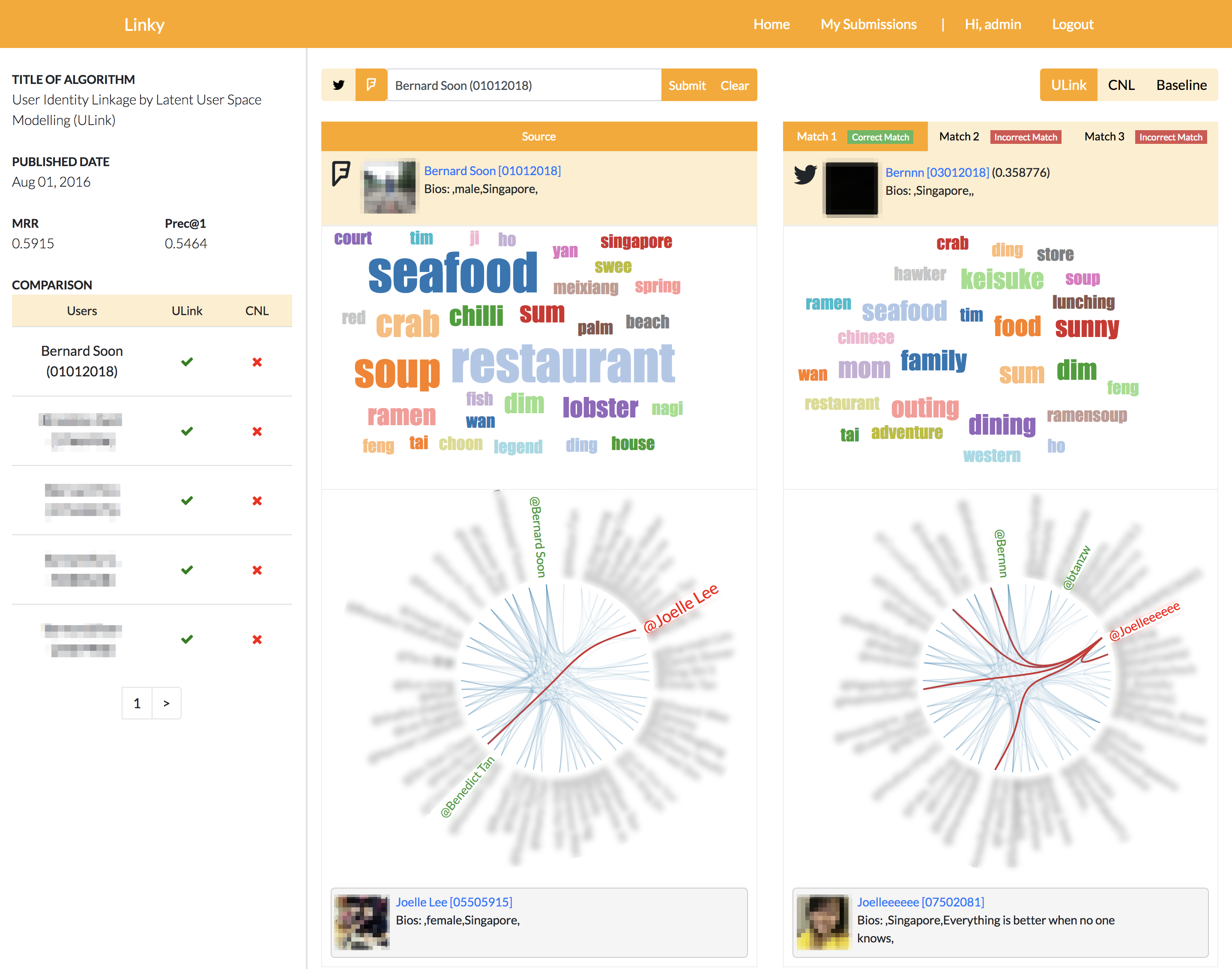}
		\caption{The \textsf{Linky} visual analytical interface. The left panel shows a list of source OSN user identities which are correctly matched by the selected user identity linkage method and not other methods. The information of the individual source and target OSN linked user identities are displayed in middle and right panel respectively.}
		\label{fig:demo}
	\end{center}
\end{figure*}

\section{Demonstration}
\label{sec:demo}
The demonstration is designed to allow users to perform two main functions: (i) exploration of similarities and differences between a user's identities in different OSNs, and (ii) comparison of results generated by different user identity linkage methods.

\textbf{Exploring user identities across multiple OSNs}. One can explore the similarities and differences between the linked user identities. Other than inspecting their profile attributes, \textsf{Linky} supports the inspection of ego networks of the linked user identities to see if they share any neighbor (or friend) which have also been linked (i.e., neighbor's identities in two OSNs has been linked to belong to the same users). The ego networks of the linked user identities' source and target OSNs are also synchronized such that if a neighbor within an $OSN_1$ has a user identity linked to a neighbor in $OSN_2$, the username of the neighbor will be highlighted in green in both $OSN_1$ and $OSN_2$ ego networks of the linked user identity. This enables visitors to observe how the matched user maintain his or her relationship across multiple OSNs. If a user maintains many overlapping friends in multiple OSNs, methods using network information should be able to return correctly linked user identities for this user. In Figure \ref{fig:demo}, Foursquare user \textit{@Bernard Soon} has a Foursquare friend  \textit{@Benedict Tan} which is linked to \textit{@btanzw}, a Twitter user which is in \textit{@Bernard Soon}'s Twitter ego network. 

One can also easily examine the relationships among the neighbor identities, i.e., beyond the links with the ego user identity, by mousing-over each neighbor identity and observing the identity's relationships highlighted in red (See Figure~\ref{fig:demo}, example neighbor \textit{@Joelle Lee}). The mouse-overed neighbor's user attributes (e.g., screen name, bios, etc.) will also be displayed below the graph. This visualization feature allows \textsf{Linky} users to observe how well-connected is a particular neighbor with the other neighbors of the user. The "closer" neighbors with more common neighbors can, therefore, be quickly identified. 

Finally, \textsf{Linky} also displays two \textit{word clouds} which summarize the content generated by the two user accounts. One can then compare the topical interests of the matched accounts. For example,  in Figure \ref{fig:demo}, Foursquare user \textit{@Bernard Soon} posted mainly food-related content in Foursquare and while he shared a mix of family and food-related topics on Twitter. 

\textbf{Comparing user identity linkage methods}. The user first selects to view the results of a particular user identity linkage methods (see Figure \ref{fig:solution}). Next, search by username can be performed to locate a user identity in either the source or target OSN. The search result will visualize present the linked user identities' profiles, content and ego networks from the two OSNs (as shown in Figure~\ref{fig:demo}). Each matched user account is also associated with a score returned by the user identity linkage method to indicate the matching confidence. For example, for a given Foursquare account with screen name \textit{@Bernard Soon}, the ULink method returns the Twitter identity with username \textit{@Bernnn} with a score of 0.358, while the CNL method returns another Twitter identity with username \textit{@bernda} with a score of 0.274.  While the two scores may not be directly comparable, we can use \textsf{Linky} to validate the results by comparing the two matched Twitter identities' user attributes (e.g. bio description), ego networks and summarized content. Intuitively, one could more easily judge which user linkage method returns the correct result. As each user linkage method generates a ranked list of matched user identities, we assume that some criteria such as \textit{top-k} or \textit{threshold} will be introduced to select the best linked user identities to be shown in \textsf{Linky}. In this demonstration, we use \textit{top-3} criteria to evaluate results from all user identity linkage methods. In the result matched target user identity panel (i.e., in the rightmost panel in Figure \ref{fig:demo}), \textsf{Linky} user can select to view and compare the top 3 matched user identities returned by the selected algorithm by toggling the tabs at the top of the panel.

\begin{figure}[h]
	\begin{center}
		\includegraphics[scale = 0.42]{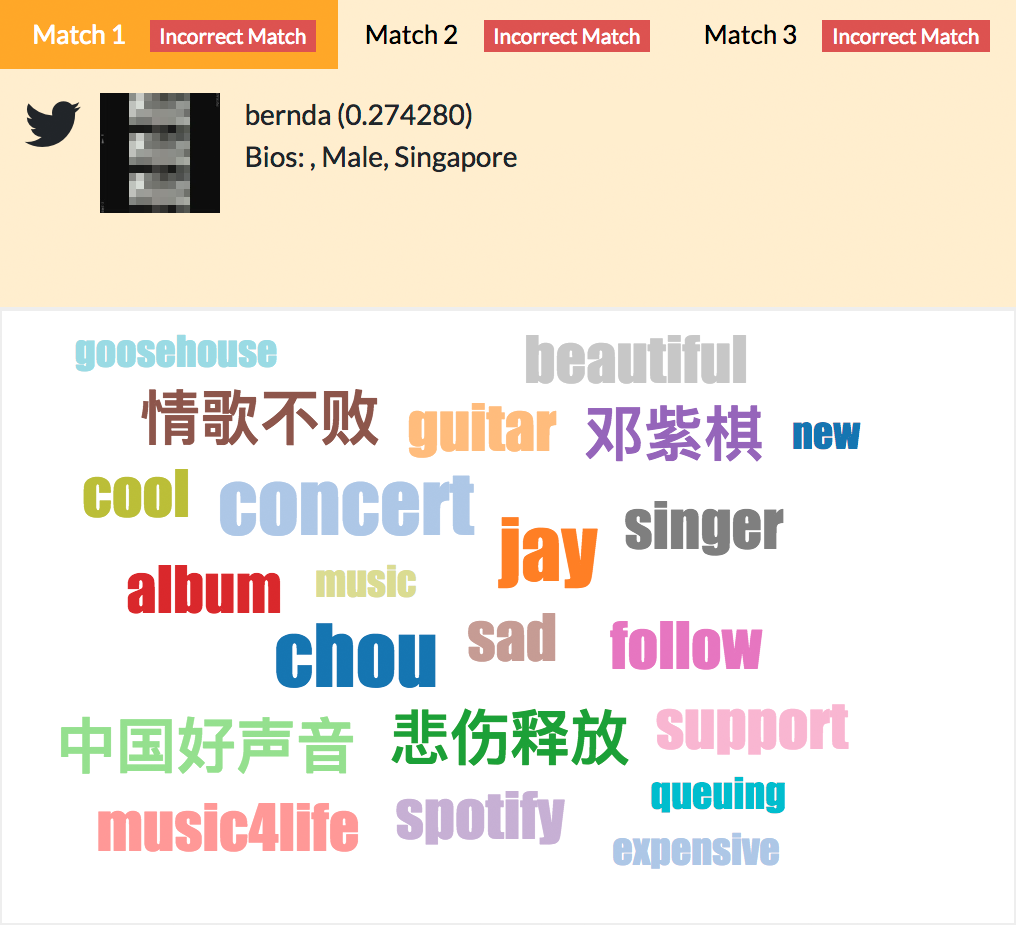}
		\caption{Content word cloud of Twitter user @bernda returned by CNL}
		\label{fig:brenda}
	\end{center}
\end{figure}

We can also empirically inspect the strengths and weaknesses of different user identity linkage methods. As shown in Figure~\ref{fig:demo}, the left panel in \textsf{Linky} visual analytical interface shows a list of user identities which are correctly linked by the selected user identity linkage method (i.e., ULink) but wrongly linked by CNL. The ground truth user identity pairs are obtained from the subset of users who have declared their corresponding OSN accounts in their short bio description (as described in Section \ref{sec:architecture}).  By exploring the profile, content and ego networks of these user identities may give us intuition on the strength of the selected user identity linkage method. For example, the Twitter user identity of Foursquare user \textit{@Bernard Soon} is correctly linked using ULink but not CNL. In the user identity linkage results of user \textit{@Bernard Soon}, we can see that the linked Twitter user identity (i.e., \textit{@Bernnn}) returned by ULink shares similar generated content with \textit{@Bernard Soon} but such observation cannot be found in the Twitter user identity (i.e., \textit{@bernda}) returned by CNL (Refer to Figure \ref{fig:brenda}). This may give us the intuition that ULink outperforms CNL in this particular result as it is able to better utilize the users' content for user identity linkage.  

\section{Conclusion}
\label{sec:conclusion}
We demonstrate \textsf{Linky}, a visual analytic tool that complements existing user identity linkage evaluation metrics to uncover micro-level insights on the user identity linkage results, and determines the strengths and weaknesses of different user identity linkage methods. \textsf{Linky} has two major use cases. Firstly, the tool enables researchers working on user identity linkage problem to analyze and compare the linked user identities generated by state-of-the-art methods at the individual user identity level. Next, the tool allows users to explore and examine the similarities and differences in users' attributes, relationship networks and generated content of the matched user identities. In the future work, we will explore other complex visualizations that could help researchers to better understand and quantify the contributions of different features in the user identity linkage methods.

\section*{Acknowledgment}
This research is supported by the National Research Foundation, Prime Minister's Office, Singapore under its International Research Centres in Singapore Funding Initiative.

\bibliography{ref.bib}
\bibliographystyle{IEEEtran}

\end{document}